\documentstyle[twocolumn,aps,epsfig]{revtex}

\begin{document}
\wideabs{
\title{\Large Dynamical properties of constrained drops}

\author{M.Ison $^1$, P.Balenzuela $^1$,A.Bonasera $^2$ and C.O.Dorso $^1$}

\address{\it $^1$ Departamento de F\'{\i}sica-Facultad de Ciencias Exactas y Naturales, Universidad de Buenos
 Aires, Pabell\'on 1 Ciudad Universitaria, 1428 Buenos Aires, Argentina}
\address{\it $^2$ INFN-Laboratorio Nazionale del Sud, Via S.Sofia 44, I-95123 Catania, 
Italy}

\maketitle
\date{\today}

\begin{abstract}
{\it In this communication we analyze the behavior of excited drops contained 
in spherical volumes. We study different properties of the dynamical systems
i.e. the maximum Lyapunov exponent $MLE$, the asymptotic distance in
momentum space $d_{\infty }$ and 
the normalized variance of the maximum fragment $NVM$. It is
shown that the constrained systems behaves as undergoing a first order phase
transition at low densities while as a second order one at high densities.
The transition from liquid-like to vapor-like behavior is signaled both by
the caloric curves, thermal response functions and the $MLE$.
The relationship between $MLE,d_{\infty }$, and the $CC$ is explored.}
\end{abstract}

\bigskip

and fluctuations.
}

\section{Introduction}

Since the advent of accelerators powerful enough to explore the behavior of
nuclear systems at intermediate energies, a new field of research has been
opened, the thermodynamics of small systems. This problem has emerged since
the first suggestion that a highly excited nuclear system might be
undergoing a second order phase transition. Such a hypotesis was brought up
by the pioneering work of the Purdue group in which a power law was fit to
the mass spectra in collisions of highly energetic protons against heavy nuclei. 
Since then many experiments have
been performed in this energy range (for a recent review see \cite{bonador}).
In order to gain understanding of the physical phenomena involved in such a
process different models have been devised. This models can be roughly
classified into two main groups, the statistical ones and the dynamical
ones. In the first case it is assumed that the highly excited nuclear system
resulting from the collision of two heavy ions is able to equilibrate in a
fixed volume (usually referred to as the freeze-out volume) and then fragments.
This fragmentation process is entirely driven by the available phase space.
This model has been quite successful in describing some aspects of the
fragmentation phenomena. In what is relevant for our present analysis we
recall that the caloric curve ($CC$), the functional relationship
between the energy and the temperature of the system, predicted in this
approach is of the ''rise-plateau-rise'' type and consequently displays a
vapor branch. The second approach, the dynamical one, assumes that the
system is properly described by a given Hamiltonian and puts no constrain in
the evolution of the excited system. In this case no equilibration has been
found. It has been shown \cite{stracha} that the resulting $CC$ is of the
''rise-plateau'' type, and no vapor branch is present. In this case the $CC$
is defined as the functional relationship between the temperature and the
energy of the system \textit{at fragmentation time}.

This two views of the process of fragmentation give
interesting insight on the properties of a fragmenting system, but differ in
the relevant issue of the degree of equilibration of the system. It is then of
primary importance to have a clear understanding of the thermodynamical and
dynamical properties of finite systems enclosed in constraining volumes,
thus having the possibility of reaching equilibrium. Some studies have been
performed in this direction. For example in \cite{gulminelli} the behavior
of lattice gases constrained in fluctuating volumes has been analyzed. Moreover 
in \cite{gross} a rather extensive analysis of
this kind of systems was performed. In \cite{campi} a phase diagram was
 built.

In this paper we focus on the analysis of drops formed by 147
Lennard Jones particles enclosed in different volumes performing a study in
 terms of the amount of energy added to the system and its
density.

The main reason for choosing this particular system is that in a series of
previous works we have performed a detailed, though still incomplete, study
of its properties when no constrain is imposed \cite{stracha,balin,enepart}.
 Moreover we have recently
given a first step towards the characterization of the behavior of
equilibrated systems \cite{cherno1}.The sizes of the constrained volumes have
been chosen such that we go from a dilute case to a very dense one. In order
to study such a system we performed extensive numerical simulations
of the Molecular Dynamics type.

This paper is structured in the following way: In section I we briefly
describe the model used in our numerical simulations.

In section II we show the calculated Caloric Curves ($CC$) for our test
system composed of 147 Lennard Jones particles as a function of its energy
for different values of the radius of the constraining volume. 

In section
III we describe the methodology we use to calculate the MLE and the $d_{\infty }$. 

In section IV we show our results. Finally conclusions are drawn.

\subsection{The Model}

Following a series of previous works in this field \cite
{stracha,balin,cherno1,cherno2,aldo} we will rely heavily on numerical
simulations of classical systems interacting via a Lennard Jones potential,
which reads:

\begin{equation}
V(r)=\left\{ 
\begin{array}{ll}
4\epsilon\left[\left(\frac{\sigma }{r}\right)^{12}-\left(\frac{\sigma}{r}%
\right)^{6}-\left(\frac{\sigma }{r_{c}}\right)^{12}+\left(\frac{\sigma}{r_{c}%
}\right)^{6}\right] & r<r_c \\ 
0 & r \ge r_c
\end{array}
\right.
\end{equation}

\smallskip We fix the cut-off radius as $r_c=3\sigma $. Energy and distance
are measured in units of the potential well ($\epsilon$) and the distance at
which the potential changes sign ($\sigma$), respectively. The unit of time
used is: $t_0=\sqrt{\sigma^{2}m/48\epsilon}$. In our numerical experiments
initial conditions were constructed using the already presented \cite
{stracha} method of cutting spherical drops composed of 147 particles out of
equilibrated, periodic, 512 particles per cell L.J. system. We choose this
kind of initialization because we consider that the resulting correlations
present in the system at initial time are the least biased ones. It is worth
mentioning at this point that the general features of the fragmentation
process do not depend on the initial state as has been shown in \cite
{cherno2} were the fragmentation of bidimensional drops via the collision
with fast aggregates of three particles was studied. A broad energy range
was considered such that the asymptotic mass spectra of the fragmented drops
displays from ''U shaped'' pattern , to exponentially decaying one.
Somewhere in between this two extremes a power law like spectra can be
found.

\smallskip

Although our system is purely classical and no direct connection with nuclear
systems can be established, one has to take into account that the main
features of the nuclear interaction (strongly repulsive at very short range
and attractive at a longer range) are present in this interaction potential.
Then it is quite plausible that the main features of L.J. systems should
appear in nuclear systems. In this respect, and of great importance for this
work, both systems present an equation of state of the same type.

Because we will be mainly interested in the behavior of constrained systems
we have to define the walls that will contain the excited drop. We have
defined the walls of our container using a very strongly 
short ranged repulsive potential (cut and shifted) defined as:
\begin{equation}
V_{w}=\left\{ 
\begin{array}{lll}
\exp \left( 1/(r-R)\right) -\exp \left( 1/(r_{c}-R)\right) & 
\Leftrightarrow & r_{c}\leq r\leq R \\ 
0 & \Leftrightarrow & r<r_{c}
\end{array}
\right. 
\end{equation}
Where $r_{c}$ defines the skin of the constraining volume.

\subsection{Caloric Curves at constant density of constrained Systems}

The Caloric Curve ($CC$) is one of the main observables in the analysis of
multifragmentation. In fact there is no agreement in the nuclear physics
community about its properties. The analysis of experimental data has given
different results. Different Thermometers have been used and the resulting $%
CC $s are of two types. On the one hand we have those that present a
''rise-plateau-rise'' shape which resemble the standard view, inherited
from classical thermodynamics, and have induced to recognize a transition
from a liquid-like to a vapor-like state. This same result has been
obtained when one adheres to statistical models to analyze the phenomena.
Because statistical models impose equilibrations in a given ''freeze-out''
volume it is natural to get this kind of behavior. On the other hand
classical molecular dynamics calculations indicate \cite
{stracha,cherno2} that when the process of fragmentation is
properly analyzed in phase space \cite{dorsorandrup} the resulting caloric
curve, in this case defined as the temperature of the system at
fragmentation time, is of the type ''rise-plateau'' and no vapor branch is
present. This behavior has been traced to the presence of a collective
motion, expansion, that behaves as a ''heat sink''. Recent experimental
results seem to confirm this view \cite{natowitz}. 

\vspace{0.5cm}

\begin{figure}
\centerline{\epsfig{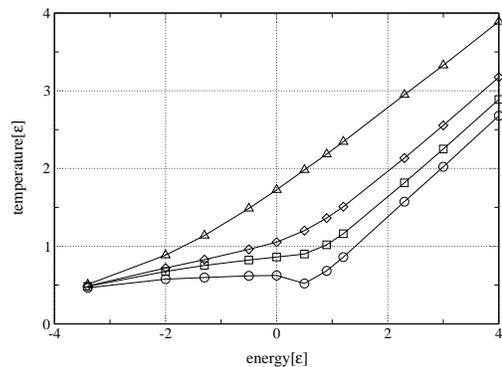}}
\caption{ In this figure we show the Caloric Curve for 147 Lennard Jones
particles for different sizes of the constraining volumes. Circles denote
the CC corresponding to a constraining volume of radius $R=15\sigma $ ,
squares for $R=8\sigma ,$ diamonds for $R=6\sigma $ and finally triangles
denote the CC for $R=4\sigma$}.
\end{figure}

In a recent work we have
shown that the effect of confining the excited drop in a given volume, i.e.
allowing the system to reach equilibrium, is the appearance of the vapor
branch. 
In this work we will extend those calculations exploring a broader range of
densities and incorporing useful dynamical quantities.

\begin{figure}
\centerline{\epsfig{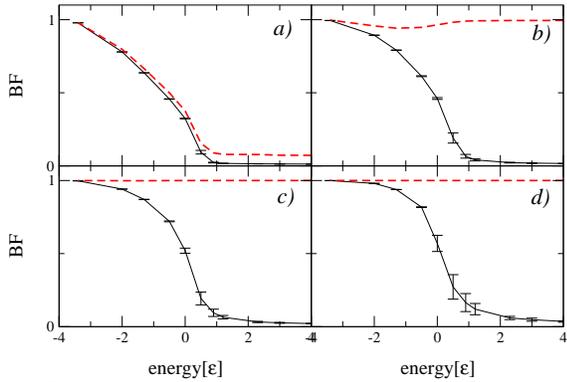}}
\caption{ In this figure we show the size of the biggest fragment, normalized
to the total size of the system, as a function of the energy deposited in
the system and for the 4 volumes considered in this work. Panel a)
corresponds to a volume of radius $R=15\sigma ,$ while b) corresponds to  $%
R=8\sigma,$ c) $R=6\sigma $ and d) $R=4\sigma .$ In each panel two curves
are drawn, full lines correspond to fragments recognized by the ECRA
formalism, while dotted lines correspond to MST analysis. It can be
immediately seen that in the dilute case ($R=15\sigma $) both descriptions
give essentially the same result. On the other hand as the system becomes
dense the MST comprises most of the mass of the system. This is the reason
why as the density is increased the system is not able to separate in
different coexisting phases in configurational space. On the other hand, in
phase space the system is fragmented irrespective of the value of the density}.
\end{figure}

From the analysis of molecular dynamics calculations of constrained systems
 \cite{cherno1} we have extracted the $CC$s that are displayed in fig.1). It
is immediate that a very interesting phenomenon takes place. The same system
at the same temperature will change its behavior as a function of the
density. In this figure it can be clearly seen that for the less dense case
a clear loop in the $CC$ is obtained. But as the density is increased this
loop disappears and is replaced by a change in the slope. Moreover at even
higher densities the $CC$ looks essentially straight and all signals of a
change in behavior are erased.

The origin of the relationship between the $CC$ and the density can be 
understood quite easily by  
analyzing the biggest cluster formed in
the system. For this purpose we use two algorithms already presented in the
literature \cite{stradormst} (see Appendix for details). Very briefly we can
say that the, improperly called, Minimum Spanning Tree (MST) algorithm looks for
clusters of interacting particles in configuration space, and completely
disregards the relative momentum. On the other hand we use the Early Cluster
Recognition Algorithm (ECRA) which seeks for the most bound partition 
 in phase space. This algorithm has allowed us to find that the
fragmentation in an expanding system takes place very early in the evolution.
In fig.2) we show the obtained biggest fragment from the analysis of
configurations according to both techniques in constrained systems (see
caption for details).

\begin{figure}
\centerline{\epsfig{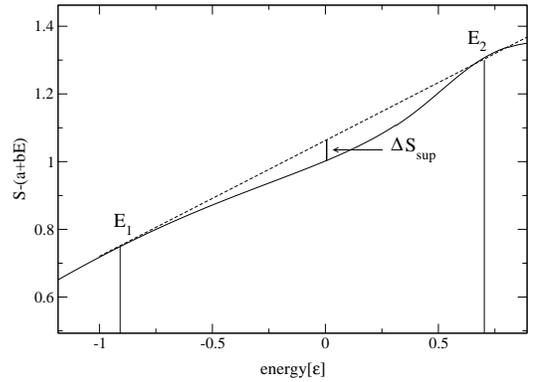}}
\caption{ In this figure we display the resulting value of the entropy as a
function of the energy for a constraining volume of radius $R=15\sigma .$
The curve displays a convex intruder between $E_1$ and $E_2$ which has been 
associated with a first
order phase transition, $\Delta S$ being the entropy lost in the formation of 
surfaces. In order to visualize the convex intruder a lineal function 
$a+b\,E$ has been substracted to the entropy. Here $a=4.3$, $b=1.4$}.
\end{figure}

The emerging picture is the following: In the case that the fragment
recognition algorithm is the MST, we see that for low densities there is
enough room in the constraining volume to allow the formation of drops i.e.
 At low densities the biggest MST fragment is a decreasing function of the
energy. But as we increase the density the biggest MST fragment comprises
most of the mass in the system. On the other hand ECRA analysis shows that
for all the densities considered the ECRA biggest fragment is a decreasing
function of the energy.

\vspace{1cm}

\begin{figure}
\centerline{\epsfig{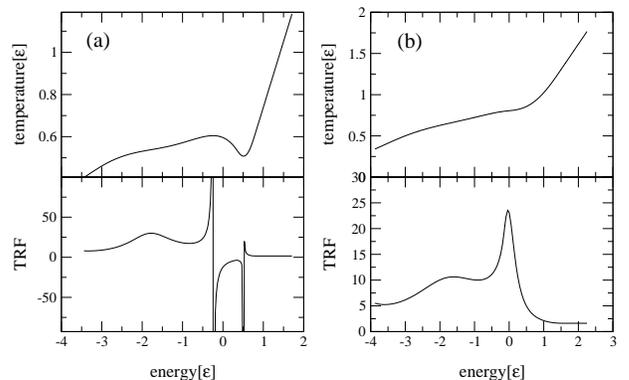}}
\caption{ Here we show the CC (upper panels) and the associated TRF (lower
panels) for two values of the volume. Left hand side correspond to $%
R=15\sigma $ (dilute case) and right hand side to $R=8\sigma $ (transition
to dense state). It can be seen that for the dilute case two poles are
present which limit the region of negative TRF. On the other hand at $%
R=8\sigma $ the TRF is always positive and displays a maximum}.
\end{figure}

Once the $CC$ is known it is easy to calculate the Entropy $S$ as
a function of the energy and the density.

\begin{equation}
S=\int \frac{dE}{T(E)} 
\end{equation}

In fig.3) we show $S$ for a dilute system (see caption for
details). It is immediate that a convex intruder appears  
which has been proposed to be a signature of a first order phase
transition in non extensive systems. (i.e. the formation of surfaces turns the 
entropy into a non-extensive function in small systems).

The next step is to calculate the behavior of the Thermal response function
of such a system.

\begin{equation}
TRF=\left( \frac{dE}{dT}\right) 
\end{equation}

In fig.4) the results of such calculation are displayed. We can see that
for low densities, as a consequence of the presence of a loop, two poles and
negative values are attained by this quantity. This has been signaled as an
evidence for a first order phase transition. It is due to the fact that
surfaces appear in the system. As the density is increased we reach a
threshold above which the $CC$ only displays a change in the slope. Then the
two poles merge in a single finite maximum (finite size effect). We can then
state that the
system goes from a first order like to a second order like behavior as a
function of the available space in configuration space.

\subsection{Maximum Lyapunov exponent and Asymptotic distance in momentum
space}

We now proceed with our analysis and focus on the dynamical aspects of
the constrained system and its relationship with the above described
thermodynamical properties.

One of the main tools to study a chaotic system is the maximum Lyapunov
exponent \cite{lyapu}, which is a measure of the sensitivity of the system to
initial conditions and also gives an idea of the velocity at which the
system explores the available phase space. Given two very close initial
conditions in phase space, the MLE, ${\hat \lambda }$, is given by the
following relation :

\begin{equation}
\lambda =\lim_{t\rightarrow\infty} \lim_{d(0)\rightarrow 0} \left[\frac{1}{t}%
ln\frac{d(t)}{d(0)}\right]. 
\end{equation}

where $d(t)$ is the distance in phase space between two trajectories ($1$
 and $2$) which
initially differs each other in a very small quantity $d_0$.

In order to calculate this quantity we must define a metric:

\begin{equation}
d_{12}(t)=\sqrt{\sum_{i=1,N}\left[ a\left( r_{1}\left( t\right) -r_{2}\left(
t\right) \right) ^{2}+b\left( p_{1}\left( t\right) -p_{2}\left( t\right)
\right) ^{2}\right] } 
\end{equation}

In this equation $a$ and $b$ are constants which take care of the units. It
has been shown that the $MLE$ are independent of the metric \cite{oseledec}.
In our case we have found it useful to take $a=0$ and $b=1/m.$ With m the mass of
the particles that according to the units defined in the model gives $b=1.$

If we calculate the distance in momentum space between nearby trajectories we find
 an exponential growth, followed by a saturation. This saturation is due to the fact
 that the available phase space is limited (The energy is a constant of motion). In 
order to handle this we calculate the $MLE$ following a method due to  
 Benettin \cite{benettin}%
. In this method,
after a time step $\tau << \tau_{sat}$, the distance $d(\tau)=d_1$ is
re-scaled to $d_0$ in the maximum growing direction and the quantity $%
ln[d_1/d_0]$ is saved. Repeating the procedure at every time step $\tau$,
the logarithmic increments ${ln[\frac{d_i}{d_{i-1}}]}$ are collected. The
MLE is defined as:

\begin{equation}
\lambda=\lim_{n \rightarrow \infty} \frac{1}{n \tau } \sum_{i=1}^n {ln \left
| \frac{d_i}{d_{i-1}} \right |}
\end{equation}

The ratio $\frac{d_i}{d_{i-1}}$ is a measure of the exponential divergence
between two initially nearby orbits along the maximum growth direction at
time $i\tau$.

Another quantity that has been recently proposed is the $d_{\infty }$ \cite{dorsobona,plj}. This
is a measure of the maximum distance in momentum space when two initially
close trajectories are followed in time. In order to calculate $d_{\infty }$
we use the same metric but no rescaling as in the Benettin method is used.
In this case the molecular dynamics evolutions needed to calculate such a
quantity are much shorter due to the fact that $d_{\infty }$ reaches a
plateau rather fast. 

\section{Results}

In what follows we show the results of our calculations.

In first place we show the $MLE$ for constrained systems. In fig.5) we show
this quantity for four densities (the same densities at which the $CCs$
are shown in fig.1)). The following features are relevant: For energies
below 0, the $MLE$ is an increasing function of the energy, following the
behavior of the $CC$. As energy is increased the behavior of the $MLEs$
changes abruptly. In the range $0\lesssim E\lesssim 1$ we find that, for the
low density case the $MLE$ displays a very pronounced loop which is in
correspondence with the loop displayed by the $CC$. On the
other hand for the next two densities a clear loop is present in the $MLE$
while the corresponding $CCs$ only show a change in the slope in this
region. Finally for the highest density considered in this work the $CC$ is
featureless in this region while the $MLE$ shows a valley.

\begin{figure}
\centerline{\epsfig{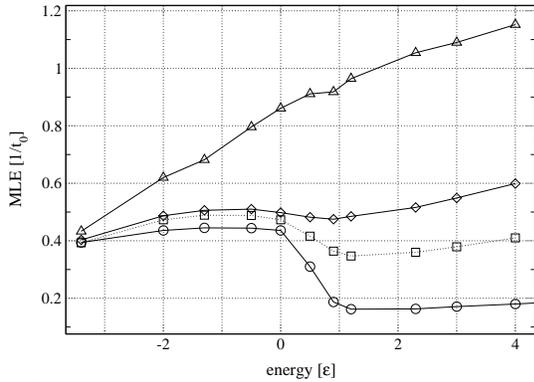}}
\caption{ In this figure we plot the $MLE$ as a function of the energy
deposited in the system for the four values of the constraining volume
considered in this work. Symbols have the same meaning as in Fig.1). It can
be seen that the $MLE$ clearly signals the transition from a liquid-like
regime to a vapor-like regime (according to ECRA analysis)}.
\end{figure}

In order to gain insight into the reasons of this behavior we have found it
useful to study the following quantities. In first place we look at the mass
distributions. When dealing with $MST$ spectra we see that two very
different behaviors appear. For $R \lesssim 15\sigma$ fragments (drops) of
different sizes appear in the volume as energy is added to the system. The
mass spectra goes from extreme U shape in the liquid like region to
exponentially decaying in the vapor like region. It is clear that as the
density is raised there will be less and less room for the system to form
non interacting drops. It is then seen that for $R \gtrsim 8\sigma$ only one $%
MST$ cluster appears in the system. At this point the $CC$
also shows a change in behavior going from a curve that displays a loop to
one that only shows a change in slope. On the other hand when the same
configurations are analyzed using the $ECFM$ , via its algorithmic
implementation : the $ECRA$ method, a different picture appears: 
 Regardless the size of the constraining walls, the system behaves as undergoing
 a phase transition, i.e. drops are formed in phase space.

We have also calculated the Normalized Variance of the size of the Maximum
fragment ($NVM$). It has been shown \cite{latora} that this magnitude will display a
maximum for a system whose mass spectra is well described by a scaling law
of the type
$n_{s} \propto s^{-\tau }f(z)$ at $f(z)=1$. In other words at
the point in which the mass distribution is a power law, i.e. a distribution
function free of scales. $NVM$ is defined as

\begin{equation}
NVM=\frac{\sigma ^{2}\left( BF\right) }{<BF>} 
\end{equation}

Where $BF$ is the normalized mass of the biggest fragment and the brackets indicate an
 ensemble averaging.
In fig.6) we show the $NVM$ for the four densities considered. It can be
readily seen that this quantity displays a maximum in the energy range in
which the $MLE$ displays a loop (or a valley for the highest density
considered).

\begin{figure}
\centerline{\epsfig{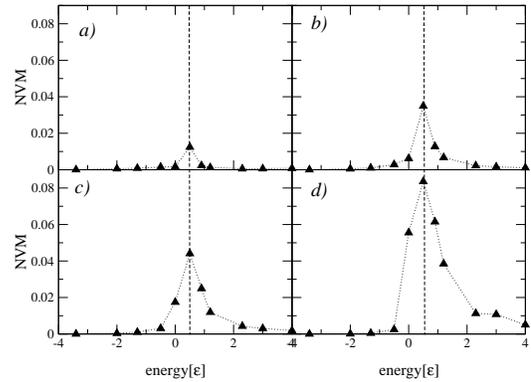}}
\caption{ Values of the $NVM$ as a function of the energy for the volumes
considered in this work. a) corresponds to $R=15\sigma ,$ b) to $R=8\sigma
,$ c) to $R=6\sigma ,$ d) to $R=4\sigma .$ Notice that it displays a clear
maximum in all cases which correspond to the region in which an anomaly is
detected in the behavior of the $MLE$}.
\end{figure}

Now we consider the behavior of the other relevant magnitude presented in
this work, i.e. $d_{\infty }.$ 
 In order
to explore its scaling properties we define $d_{\infty }^{n}$ which stands
for the normalized asymptotic distance

\begin{equation}
d_{\infty }^{n}(t)=\left( \frac{\sum_{i=1}^{N}\left[ \left(
p_{1}(t)-p_{2}(t)\right) ^{2}\right] _{i}}{\frac{1}{2}\sum_{i=1}^{N}\left[
p_{1}(t)\right] _{i}^{2}}\right) ^{\frac{1}{2}} 
\end{equation}

In other words

\begin{equation}
d_{\infty }^{n}(t)=\frac{d(t)}{\sqrt{K}} 
\end{equation}

i.e., the distance in momentum space normalized to the square root of the
kinetic energy.

\vspace{1cm}

\begin{figure}
\centerline{\epsfig{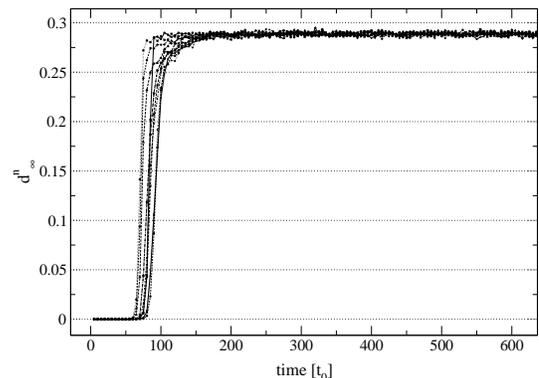}}
\caption{ $d_{\infty }$ scaled with the square root of the kinetic energy for 
a volume of radius $R=8\sigma .$ Different curves correspond to different
energies in the range $-3.4\leq E\leq 4.0$ Notice that after a short time
all the curves collapse into a single one}.
\end{figure}

In fig.7) we show the result of such an analysis for $R =8\sigma$ and in fig.8)
the same but at $R =15\sigma$ . In both cases it is seen that the asymptotic
distance in momentum space scales as the square root of the kinetic
energy (and then as the square root of the temperature) of the system as it was 
conjectured in \cite{dorsobona}.

\vspace{0.5cm}

\begin{figure}
\centerline{\epsfig{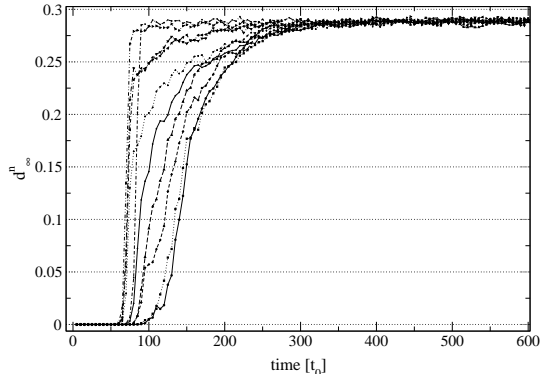}}
\caption{ Same as Fig 7.) but for a volume of radius $R=15\sigma .$ Notice
that, as before, after a still short time, but larger than in the case $%
R=8\sigma ,$ all the curves collapse into a single one}.
\end{figure}

Finally in fig.9) we show the $d_{\infty }$ as a function of the energy
for two densities. It can be seen that its behavior clearly resembles the
corresponding one for the $CC$, thus this quantity could be used instead of T
 if the latter is not reliably measurable.

\vspace{0.5cm}

\begin{figure}
\centerline{\epsfig{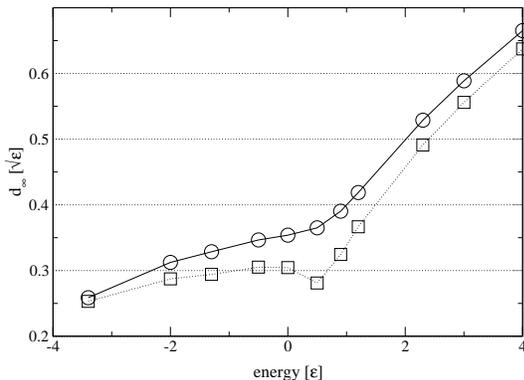}}
\caption{ In this figure we show $d_{\infty }$ as a function of the energy for
volumes corresponding to $R=15\sigma $ (squares) and $R=8\sigma $ (circles).
Upon comparison of these curves with the ones corresponding to the CC
(fig.1) we can see that both are quite similar}.
\end{figure}

\section{Conclusions}

The main results of our calculations can be summarized in the following:

a) The $CC$ is strongly dependent on the density of our system. As
the density is increased the behavior of this function goes from displaying
a loop (which we refer to as first order like behavior) to one that only
presents a change in the slope (in this case we talk about a second order
like behavior). In the first case the $TRF$ displays two poles, which
converge into a single maximum as the density surpasses a given threshold $%
(R \sim 8.7\sigma \Rightarrow \rho \sim 0.05\sigma^{-3}).$ Three regions can then be recognized, The liquid like, the
transition region and the vapor like region.

b) The $MLE$ is quite sensitive to the transition from liquid like to vapor
like states of the finite constrained system. In all cases, even when the $%
CC $ is almost featureless a signal is detected in the $MLE.$ This signal
has been found to be significant by studying other quantities like the $NVM.$

c)The $d_{\infty }$ has been shown to scale like $\sqrt{K}$ and then as
the $CC$, giving useful insight into the amount of Phase Space visited by the
system.

\section{Acknowledgment}

This work was partially supported by the University of Buenos Aires (UBA) via grant
 No tw98, and CONICET via grant No 4436/96.

C.O.Dorso is a member of the carrera del investigador (CONICET), P.Balenzuela is a
fellow of the Conicet, M.Ison is a fellow of UBA.

\section{Appendix}

In previous papers the main fragment recognition algorithms currently in use 
have been fully analyzed \cite{stradormst}. The simplest definition of 
cluster is basically: a group of particles that are close to each other and 
far away from the rest. The fragment recognition method known as minimum 
spanning tree (MST) is based on the last idea (I). In this
approach a cluster is defined in the following way: given a set of 
particles $i,j,k,...$, they belong to a cluster $C$ if :

\begin{equation}
\forall \hspace{0.2cm} i\in C\hspace{0.2cm} ,\exists \hspace{0.2cm} j\in C
\hspace{0.2cm} /\hspace{0.2cm} \left| {\bf r}_i-{\bf r}_j\right| 
\leq R_{cl}
\end{equation}

where ${\bf r}_i$ and ${\bf r}_j$ denote the positions of the particles and
$R_{cl}$ is a parameter usually referred to as clusterization radius, and
is usually related to the range of the interaction potential. In our
calculations we took $R_{cl} = 3 \sigma$.

On the other hand, the early cluster formation model (ECFM) \cite{dorsorandrup}, is
based on the next definition: clusters are those that define the most bound 
partition of the system, i.e. the partition (defined by the set of clusters 
$\{C_i\}$) that minimizes the sum of the energies of each fragment according 
to:

\begin{equation}
E_{\left\{ C_i\right\} }=\sum_i\left[ \sum_{j\in C_i}K_j^{cm}+\sum_{j,k\in
C_i}V_{j,k}\right]
\end{equation}
where the first sum is over the clusters of the partition, and $K_j^{cm}$ is
the kinetic energy of particle $j$ measured in the center of mass frame of
the cluster which contains particle $j$. The algorithm (early cluster
recognition algorithm, ECRA) devised to achieve this goal is based on an 
optimization procedure in the spirit of simulated annealing \cite{dorsorandrup}.

\end{document}